\documentclass{jfm}
\usepackage{graphicx}
\usepackage{epstopdf, epsfig}
\usepackage{natbib}
\usepackage{amsmath}
\usepackage{mathtools}
\usepackage{cancel} 
\usepackage{hyperref}
\hypersetup{
    colorlinks=true,
    linkcolor=blue,
    citecolor=blue
    }
\usepackage{xcolor}
\usepackage{float}

\newcommand\norm[1]{\left\lVert#1\right\rVert}

\shorttitle{Causality in a square cylinder}
\shortauthor{{\'A}. {Mart{\'\i}nez-S{\'a}nchez} et al.}

\title{Causality analysis of large-scale structures in the flow around a wall-mounted square cylinder}

\author{{\'A}lvaro {Mart{\'\i}nez-S{\'a}nchez}\aff{1},
  Esteban L{\'o}pez\aff{2},
  Soledad {Le Clainche}\aff{3},
  Adri{\'a}n {Lozano-Dur{\'a}n}\aff{4},
  Ankit Srivastava\aff{2} \and
  Ricardo Vinuesa\aff{1}\corresp{\email{rvinuesa@mech.kth.se}}}

\affiliation{\aff{1}FLOW, Engineering Mechanics, KTH Royal Institute of Technology, Stockholm, Sweden
\aff{2} Dept. Mechanical, Materials, and Aerospace Engineering, Illinois Institute of Technology, Chicago, IL, 60616, USA
\aff{3} School of Aerospace Engineering, Universidad Politécnica de Madrid, Madrid, E-28040, Spain
\aff{4} Department of Aeronautics and Astronautics, Massachusetts Institute of Technology, Cambridge, MA, 02139, USA
}

\begin{document}

\maketitle

\begin{abstract}
The aim of this work is to analyse the formation mechanisms of large-scale coherent structures in the flow around a wall-mounted square cylinder, due to their impact on pollutant transport within cities. To this end, we assess causal relations between the modes of a reduced-order model obtained by applying proper-orthogonal decomposition to high-fidelity-simulation data of the flow case under study. The causal relations are identified using conditional transfer entropy, which is an information-theoretical quantity that estimates the amount of information contained in the past of one variable about another. This allows for an understanding of the origins and evolution of different phenomena in the flow, with the aim of identifying the modes responsible for the formation of the main vortical structures. Our approach unveils that vortex-breaker modes are the most causal modes, in particular, over higher-order modes, and no significant causal relationships were found for vortex-generator modes. We validate this technique by determining the causal relations present in the nine-equation model of near-wall turbulence developed by Moehlis \textit{et al.} (\textit{New J. Phys}, vol.~6, 2004, p.~56), which are in good agreement with literature results for turbulent channel flows.
\end{abstract}


\section{Introduction}
With the rise of population in urban areas, understanding how pollutants remain trapped within metropolitan regions is of increasing importance. Recently reported as the largest environmental health risk in Europe~\citep{Lelieveld2019}, air pollution is a major cause of premature deaths and disease. Therefore, predictive models for air-quality control are relevant to provide protection from excessive pollutant concentrations. The essential need to address sustainable development from an urban perspective is enshrined in the 2030 Agenda~\citep{Agenda2030} through the sustainable development goals (SDGs) 11 and 13, on sustainable cities and climate action, respectively. However, the available models are unable to provide the required spatio-temporal accuracy to reproduce the pollutant-dispersion patterns within cities~\citep{Torres2021}. Improved prediction and assessment techniques are urgently needed to address these issues and promote urban sustainability in the near future~\citep{Vinuesa2015}. In this study, we propose an information-theoretic analysis through causality metrics of a reduced-order model (ROM) of the flow around a wall-mounted square cylinder to gain further insight into the underlying mechanisms defining the flow, shedding light into new possibilities for future urban-flow control research. Note that although the considered flow case is significantly simpler than that in an urban environment, the focus here is in the arch vortices, which are present in the analysed flow and have an important role in pollutant transport in cities~\citep{Monnier2018,Lazpita2022}.

The flow around this type of environments is generally found to be turbulent. Due to the wide range of spatio-temporal features present in such a high-dimensional non-linear chaotic system, these flows are challenging to analyse. However, the presence of similar flow characteristics across a plethora of fluid flows has revealed the presence of dominant processes that constitute the basis of various types of flows. Modal-decomposition techniques offer the possibility to analyse non-linear and chaotic dynamics and create reduced-order models (ROMs) by defining a low-dimensional coordinate system for capturing dominant flow characteristics. Proper-orthogonal decomposition (POD)~\citep{Lumley1967} and dynamic-mode decomposition (DMD)~\citep{dmd2009, Schmid2010} are two modal-decomposition methods based on linear algebra that have been widely used to extract the dominant spatio-temporal features in fluid flows. Balanced POD (BPOD)~\citep{bpod2005}, spectral POD (SPOD)~\citep{spod2018}, higher-order DMD (HODMD)~\citep{hodmd2017} and spatio-temporal Koopman decomposition (STKD)~\citep{stkd2018} are several successful variants of POD and DMD for analysis of turbulent flows. These techniques have been assessed in the context of simplified urban flows in the recent works of \cite{Lazpita2022} and \cite{martinez2022}. They analysed the near-wake flow of finite square cylinders, which can be described as a combination of four main vortices~\citep{Hunt1978,wang_zhou_2009}: the tip (or roof) vortex, the spanwise vortex, the base vortex (a streamwise vortex formed near the cylinder base with close interaction with the wake) and the horseshoe vortex, which forms around the obstacle. The well-known arch vortex can be then described as a combination of the first three~\citep{sakamoto1983,tanaka1999,wang2006,wang_zhou_2009}, resulting in a vortical structure forming on the leeward side of a wall-mounted obstacle that consists of two legs and one roof. The flow rotates in the spanwise direction in the former vortex features and in the wall-normal direction in the latter one. This vortex has been extensively studied by the fluid-mechanics community both experimentally~\citep{Hunt1978,Oke1988,Becker2002,AbuOmar2008,Zajic2011,Kawai2012,zhu2017,Monnier2018} and numerically~\citep{Sohankar1999,Saha2003,Vinuesa2015,Amor2020,Torres2021}, due to its implications in urban-environment phenomena, \textit{i.e.} pollutant dispersion, air quality, heat propagation and impact on pedestrian comfort~\citep{Oke1988}. 

The flow around these wall-mounted obstacles is also strongly three-dimensional. As the flow initially encounters the obstacle, a recirculation bubble formed on the windward side of the cylinder induces an adverse pressure gradient that thickens the incoming boundary layer, which then produces a shear layer around the obstacle~\citep{Becker2002,wang_zhou_2009}. Simultaneously, a horseshoe vortex progressively gets wider around the two sides of the cylinder, a fact that accelerates the flow close to the obstacle due to the favourable pressure gradient induced by the geometry~\citep{Hunt1978}. A separated wake is then formed downstream the obstacle with a self-sustained oscillation process and a downward motion from the top of the obstacle, which is responsible for the widening of the wake~\citep{Vinuesa2015}. 

The topology of the near-wake flow consists of free-end downwash flow, spanwise shear flow and upwash flow from the wall, which relate to the tip, base and spanwise vortices~\citep{wang_zhou_2009}. The formation of the arch vortex is produced as a result of the previous flow features and their connection or bridge near the free end, is closely related to the symmetric shedding modes, which induce an arch-type structure even on the instantaneous field~\citep{zhu2017}. Using various modal-decomposition methods, \cite{Lazpita2022} assessed the characteristics of the arch-vortex and documented the generation and destruction mechanisms of this vortex based on the resulting spatio-temporal modes, which were classified as vortex-generator and vortex-breaker modes, respectively. As a result, they suggested that the arch vortex might be connected with the dispersion of pollutants in urban environments, where its generation leads to an increase in their concentration.

Here, we focus on the previous classification to develop a reduced-order model applying POD on a very simplified urban-flow database consisting of a single building-like obstacle.
The principle of causal inference, a core idea in many scientific disciplines but rather scarce in the field of fluid mechanics, is then used to further analyse the causal interaction of the resulting modes. Since the temporal evolution associated with the aforementioned modes is typically known, the quantification of causality among temporal signals has drawn the most attention. Time correlation between a pair of signals can usually provide a simplified approach for the quantification of causality. However, this method lacks the directionality and asymmetry which is required to estimate the causes and the effects of a given set of events~\citep{beebee2009}. The notion of casuality can be tracked back to the work of \cite{wiener1956} and was first quantified by \cite{granger1969} as a statistical test for evaluating the ability of one time series to predict another. Nevertheless, standard Granger-causality tests originally assumed a functional form in the relationship among the causes and the effects that were implemented by fitting linear auto-regressive models~\citep{wiener1956,granger1969}, which is not the ideal coupling when dealing with strongly-nonlinear systems~\citep{barnett2009}. To tackle this issue, recent works are focusing on an information-theoretical framework for the estimation of causality, namely transfer entropy~\citep{schreiber2000} and information flow~\citep{nichols2013}. Those quantities require from a method to assess the conditional dependency of the variables~\citep{shannon1948}, which is computationally expensive and involves long time-series~\citep{hlavackova2011}. However, recent progress in entropy estimators using discrete and insufficient databases has made the use of transfer entropy computationally feasible~\citep{kozachenko1987,kraskov2004,gencaga2015}. Taking everything into account, identifying cause-effect interactions between events or variables remains an ongoing challenge. We used transfer entropy as a metric to quantify causality in this case, but appropriate metrics to capture causal relationships between quantities remain to be established, particularly given the huge number of parameters whose configuration is critical in the observed causation.

Some representative examples in which the previous information-theoretical approaches were applied in turbulent flows are the works of \cite{cerbus2013}, \cite{tissot2014}, \cite{liang2016} and \cite{lozano2020}. Furthermore, a much improved and rigorous definition of causality is provided in the recent work of \cite{LozanoDuran2022a} using Shannon entropy as the main quantity of interest. In this project, we focus on the causal relations reported by \cite{lozano2020} of energetic eddies in wall-bounded turbulence. These results are compared with the causal relations obtained through transfer-entropy estimators for the low-dimensional model for turbulent shear flows developed by \cite{moehlis2004}. The aim of this is to validate the proposed method with causal interactions we expect to see in advance. Our ultimate goal is to quantify the causality among the large-scale structures driving the flow dynamics in urban environments, with the purpose of identifying the modes responsible for the creation of the arch vortex and understanding the origins of various phenomena in the flow.

The paper is organised as follows: the methodology for quantifying causal interactions among signals is presented in \S\,\ref{sec: Methods}. The causal relations present in the nine-mode model for turbulent shear flows are discussed in \S\,\ref{sec: Moehlis}. The description of the numerical simulations carried out to obtain the urban-flow database is presented in \S\,\ref{sec: Numerical simulations} together with a summary of the main flow mechanisms. The reduced-order model built upon the previous database is depicted in \S\,\ref{sec: ROM} and the causal relations among their corresponding modes are discussed in \S\,\ref{sec: Results}. Finally, the conclusions of our study are presented in \S\,\ref{sec: Conclusions}.

\section{Methods in causality}\label{sec: Methods}
We use the framework provided by information theory~\citep{shannon1948} to quantify causality among different temporal signals. The central quantity for causal assessment of the signals is the Shannon entropy, which is defined as:
\begin{equation}
    H\left(X\right) = - \sum_{x\in\mathcal{X}} P_X\left(x\right) \log P_X\left(x\right),
\end{equation}
where $X$ is the random discrete-valued variable under consideration, $\mathcal{X}$ represents its associated support set and $P_X\left(x\right)$ denotes its probability density function (pdf). This quantity measures the amount of uncertainty present in variable $X$. Using the same principle, the amount of randomness in a given pair of variables $\left(X,Y\right)$ can be quantified using the joint entropy, namely:
\begin{equation}
    H\left(X,Y\right) = - \sum_{x\in\mathcal{X}}\sum_{y\in\mathcal{Y}} P_{XY}\left(x,y\right) \log P_{XY}\left(x,y\right),
\end{equation}
where $Y$ represents another random discrete-valued variable and $P_{XY}$ is the joint pdf between $X$ and $Y$. The joint entropy is useful to estimate the amount of uncertainty on $Y$ remaining after having observed $X$, which is defined as conditional entropy,
\begin{equation}
    H\left(Y\vert X\right) =  H\left(X,Y\right) -  H\left(X\right).
\end{equation}
Within this framework, we define causality from $X$ to $Y$ as the decrease in uncertainty of $Y$ knowing the past state of $X$. This is formulated through the principle of transfer entropy~\citep{schreiber2000}, which exploits the time-asymmetry of causation (the cause always precedes the effect) by using the definition of conditional entropy, \textit{i.e.}:
\begin{equation}\label{eq: Transfer Entropy}
    T_{X\rightarrow Y}(\Delta t) = H\left(Y_{t+\Delta t} \vert Y_t\right) - H\left(Y_{t+\Delta t} \vert X_t,Y_t\right),
\end{equation}
which can be expressed in terms of Shannon entropies as:
\begin{equation}\label{eq: Transfer Entropy 2}
    T_{X\rightarrow Y}(\Delta t) = H\left(Y_{t+\Delta t}, Y_t\right) - H\left(Y_t\right) - H\left(Y_{t+\Delta t}, X_t,Y_t\right) + H\left(Y_{t}, X_t\right),
\end{equation}
where $Y_{t+\Delta t}$ represents a forwarded time-shifted version of $Y$ with lag $\Delta t$ relative to the past time series $X_t$ and $Y_t$. Therefore, it can be stated that $X$ does not cause $Y$ if and only if $H\left(Y_{t+\Delta t} \vert Y_t\right)=H\left(Y_{t+\Delta t} \vert X_t,Y_t\right)$, \textit{i.e.} when $T_{X\rightarrow Y} = 0$. This is considered as an important tool to analyse the causal relationships in nonlinear systems~\citep{hlavackova2011}. An important property of transfer entropy when compared to classical time-correlation approaches~\citep{jimenez2013} is the assymmetry of measurements, \textit{i.e.} $T_{X\rightarrow Y}\neq T_{Y\rightarrow X}$, which allows to quantify the directional coupling between systems. One can interpret this quantity as a measure of the dominant direction of the information flow, which indicates which variable provides more predictive information about the other variable~\citep{michalowicz2013}.

Due to the discrete nature of the signals, the computation of (\ref{eq: Transfer Entropy 2}) is numerically performed through estimations of the pdf of each signal and their corresponding entropy values using the $k$-nearest-neighbour entropy estimator. This method, introduced by \cite{kozachenko1987}, yields an entropy estimation that can be written as:
\begin{equation}
    \hat{H}\left(X\right) = \psi(N) - \psi(k) + \log c_d + \frac{d}{N} \sum_{i=1}^N \log \varepsilon(i),
\end{equation}
where $N$ is the number of finite samples and $\psi(\cdot)$ represents the digamma function. The parameter $d$ represents the dimension of $x$ and $c_d$ is an expression that depends on the type of norm used to calculate the distances, which represents the volume of a $d$-dimensional unit-ball (for $L_\infty$-norm, $c_d = 1$). Finally, $\varepsilon(i)$ is the distance of the $i^{th}$ sample to its $k^{th}$ neighbour. The reader is referred to \cite{kozachenko1987} and \cite{kraskov2004} for a more detailed discussion of the previous entropy estimator.

In the present work, a set of $n$ time-varying signals is analysed such that it can be arranged into an $n$-component vector defined by:
\begin{equation} \label{eq: Vector}
    \boldsymbol{\mathcal{V}}(t) = \left[\mathcal{V}_1(t),\mathcal{V}_2(t),\dots,\mathcal{V}_n(t)\right].
\end{equation}
Using this nomenclature, the transfer entropy in (\ref{eq: Transfer Entropy}) can be defined with $X=\mathcal{V}_i$ and $Y=\mathcal{V}_j$ as:
\begin{equation}\label{eq: Transfer Entropy Vector}
    T_{i\rightarrow j}(\Delta t) = H\left(\mathcal{V}_j(t+\Delta t) \vert \boldsymbol{\mathcal{V}}^{\cancel{i}}(t)\right) - H\left(\mathcal{V}_j(t+\Delta t) \vert \boldsymbol{\mathcal{V}}(t)\right),
\end{equation}
where $\boldsymbol{\mathcal{V}}^{\cancel{i}}$ represents the vector $\boldsymbol{\mathcal{V}}$ but without the component $i$. This definition was introduced in \cite{lozano2020}. It generalized the definition of \cite{schreiber2000} to multiple variables, which results in a causal map with the cross-induced cause-and-effect interactions between each signal, where the terms $T_{i\rightarrow i}$ are set to zero. Furthermore, to assess these interactions, we normalise every causal effect $T_{i\rightarrow j}$ using the $L_\infty$-norm.

Apart from the measurement-asymmetry feature, another relevant property of transfer entropy is that it only accounts for direct causal interactions excluding intermediate ones, \textit{i.e.} if $\mathcal{V}_i \rightarrow \mathcal{V}_j$ and $\mathcal{V}_i \rightarrow \mathcal{V}_k$ are unique causal relations, there is no cause interaction $\mathcal{V}_j \rightarrow \mathcal{V}_k$~\citep{duan2013}. Furthermore, transfer entropy is invariant to transformation of the signals since it is exclusively based on their associated pdfs~\citep{kaiser2002}.

\section{A low-dimensional model of the near-wall cycle of turbulence\label{sec: Moehlis}}

In this section, we analyse the casual relations present in a low-dimensional model for turbulent shear flows developed by \cite{moehlis2004}. The aim of this is to verify the ability of the proposed method to identify causal interactions we expect to see \textit{a priori}. In particular, the causal relationships identified between the modes of the low-dimensional model of the near-wall cycle of turbulence are contrasted with literature results for turbulent channel flows~\citep{lozano2020}. The model is based on Fourier modes and describes the flow between two free-slip walls subjected to a sinusoidal body force. The essential elements of the model are nine modes $\boldsymbol{\rm{v}}_j(\boldsymbol{\rm{x}})$, where $\boldsymbol{\rm{v}}_1$ represents the basic mean velocity profile; $\boldsymbol{\rm{v}}_2$, the streaks; $\boldsymbol{\rm{v}}_3$, the downstream vortex; $\boldsymbol{\rm{v}}_4$ and $\boldsymbol{\rm{v}}_5$, the spanwise flows; $\boldsymbol{\rm{v}}_6$ and $\boldsymbol{\rm{v}}_7$, the normal vortex modes; $\boldsymbol{\rm{v}}_8$, a three-dimensional (3D) mode and $\boldsymbol{\rm{v}}_9$, the modification of the basic profile. All but one of the modes were already introduced by the eight-mode model proposed by \cite{waleffe1997} for sinusoidal shear-flow turbulence with some additional couplings between modes included in the case employed in this work. An extra mode is also introduced in the model of \cite{moehlis2004} to account for the modification of the structure of the mean velocity profile, over which turbulent fluctuations are known to have a significant impact. The instantaneous velocity fields can be obtained by superposing the previous nine modes as:
\begin{equation}
    \boldsymbol{\rm{u}}(\boldsymbol{\rm{x}},t) \coloneqq \sum_{j=1}^9 a_j(t) \boldsymbol{\rm{v}}_j(\boldsymbol{\rm{x}}),
\end{equation}
where the spatial coordinates are denoted by $\boldsymbol{\rm{x}}$ and $t$ represents the time. The Galerkin projection of the Navier--Stokes equations can be then applied onto this subspace to obtain a system of nine ordinary differential equations (ODEs) in time. Each ODE term exhibits a linear term, several non-linear terms (including the interactions between modes under the shape $q_k(t) = a_i(t) a_j(t)$) and a constant. Hence, they can be generally written as:
\begin{equation}
    \frac{d \boldsymbol{\rm{a}}(t)}{dt} = \boldsymbol{L} \boldsymbol{\rm{a}}(t) + \boldsymbol{N} \boldsymbol{\rm{q}}(t) + \boldsymbol{\rm{c}},
\end{equation}
where $\boldsymbol{\rm{a}}\in \mathbb{R}^n$ represents the vector of mode amplitudes, $\boldsymbol{\rm{q}}\in \mathbb{R}^m$ is the vector of nonlinear processes, $\boldsymbol{L}\in \mathbb{R}^{n\times n}$ and $\boldsymbol{N}\in \mathbb{R}^{m\times n}$ are the matrices of coefficients for the linear and nonlinear terms, respectively, and $\boldsymbol{c}\in \mathbb{R}^{n}$ is the vector of constants. {The Reynolds number is defined as a function of the channel full height $2\ell$ and the laminar velocity $U_0$ at a distance of $\ell/2$ from the top wall. The model used here corresponds to $Re = 400$, where $\ell$ and $U_0$ are used as length and velocity scales, respectively. The domain size is $L_x=4\pi$, $L_y=2$ and $L_z=2\pi$, where $x$, $y$ and $z$ denote the streamwise, wall-normal and spanwise directions, respectively.} Over this domain, the ODE model was used to obtain 600 sets of time series of the nine amplitudes, each with a time span of $4{,}000$ time units and a time step of $0.01$ time units. These sets of time series are the result of introducing a random perturbation to $a_4$.

\begin{figure}
  \centerline{\includegraphics[width=0.55\textwidth]{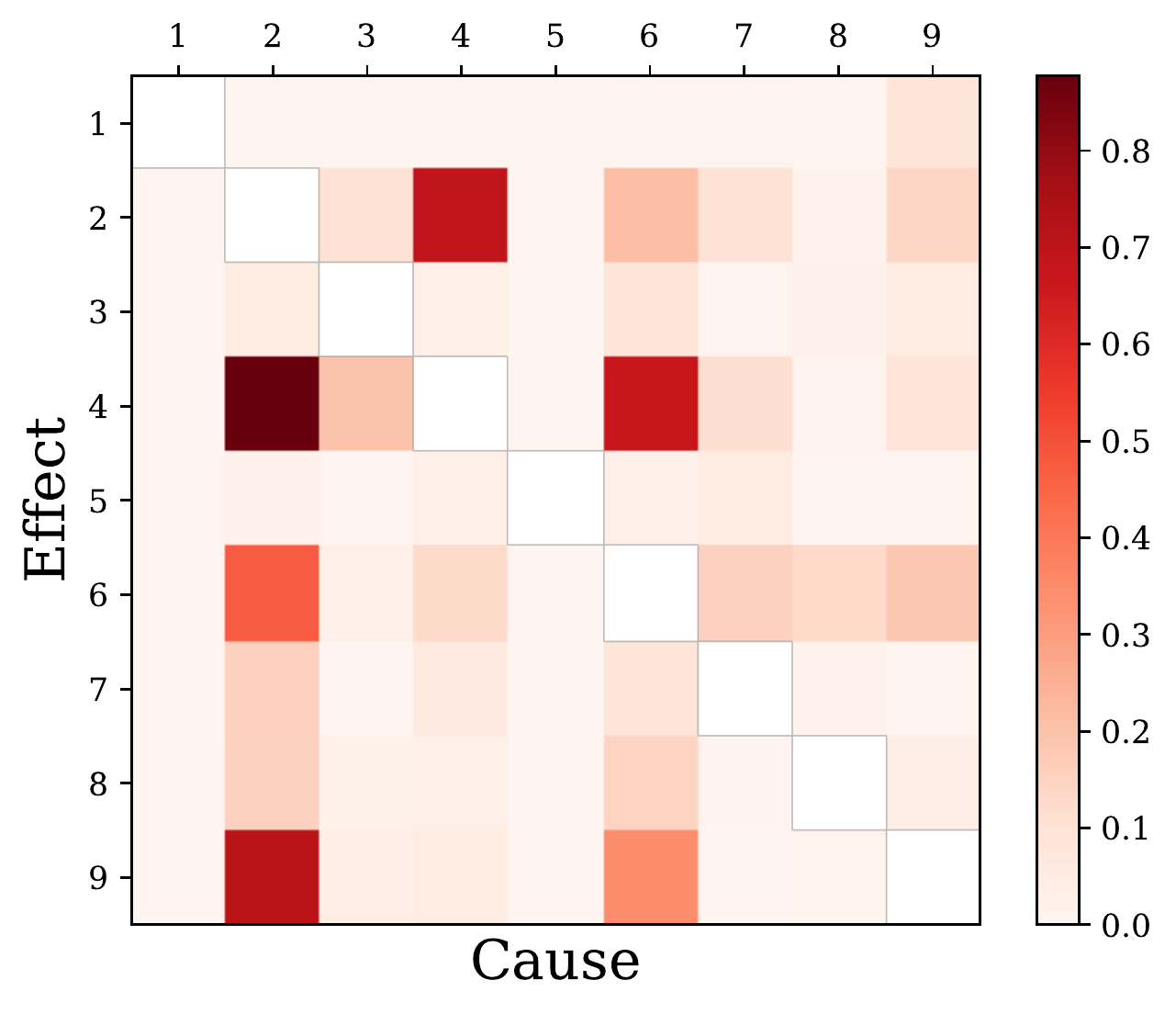}}
  \caption{Causal map for the low-dimensional model for turbulent shear flows proposed by \cite{moehlis2004}. Redscale colours denote causality magnitude normalised using $L_\infty$-norm. Modes are numbered from 1 to 9 and represent the basic profile, streaks, downstream vortex, spanwise flows, normal vortex modes, three-dimensional (3D) mode and modification of basic profile, respectively. The map is the result of the averaging of 600 time series sets with a time lag corresponding to one-snapshot lag, \textit{i.e.} $\Delta t = 0.01$.}
\label{fig:MoehlisCausality}
\end{figure}

The time-series evolution for all nine amplitudes in the model are then used to determine the casual interactions between each of the modes. The key results of this work are shown in figure \ref{fig:MoehlisCausality}, which contains the causal relations among the nine modes. The transfer entropy in (\ref{eq: Transfer Entropy Vector}) was estimated using a time lag of $\Delta t= 0.01$ and a nearest-neighbour parameter of $k=4$. Whereas the latter has been demonstrated to produce consistent results keeping the computational cost at an optimised level~\citep{kraskov2004}, varied causal interactions may be derived for different values of $\Delta t$. In the present example, time lags in the range $\Delta t = \left[0.01,0.1\right]$ were tested and no significant discrepancies were observed. A more detailed discussion of the impact of this parameter on causation is provided in \S\,\ref{sec: Results}.

The map should be read as causative variables in the horizontal axis versus the corresponding effects on the vertical axis. In a single visual, several flow mechanisms can be identified. The casual connections $a_2\rightarrow a_6$ and $a_2\rightarrow a_9$ represent how the streaks have an effect on the streamwise vortex modes and the modification of the basic profile. This causal interaction is consistent with the lift-up mechanism, where the streak amplitude is amplified through the wall-normal momentum transport~\citep{orr1907}. The causality $a_6\rightarrow a_4$ is then associated with the generation of rolls, motivated by the influence of normal-vortex modes on the spanwise flow. The most noticeable link arises from the cause-and-effect interaction $a_2\leftrightarrow a_4$, which results in an instability of the mean flow in the spanwise direction. All of these causal relations are analogous to those reported by \cite{lozano2020} in turbulent channel flow. Therefore, these findings suggest that, using the nearest-neighbours-entropy estimator to quantify transfer entropy, it is possible to extract the most relevant causal interactions between modes of a highly nonlinear system. The following sections will focus on the application of this entropy estimator to the flow around a wall-mounted obstacle.

\section{Numerical simulations and flow description\label{sec: Numerical simulations}}

\begin{figure}
  \centerline{\includegraphics[width=0.8\textwidth]{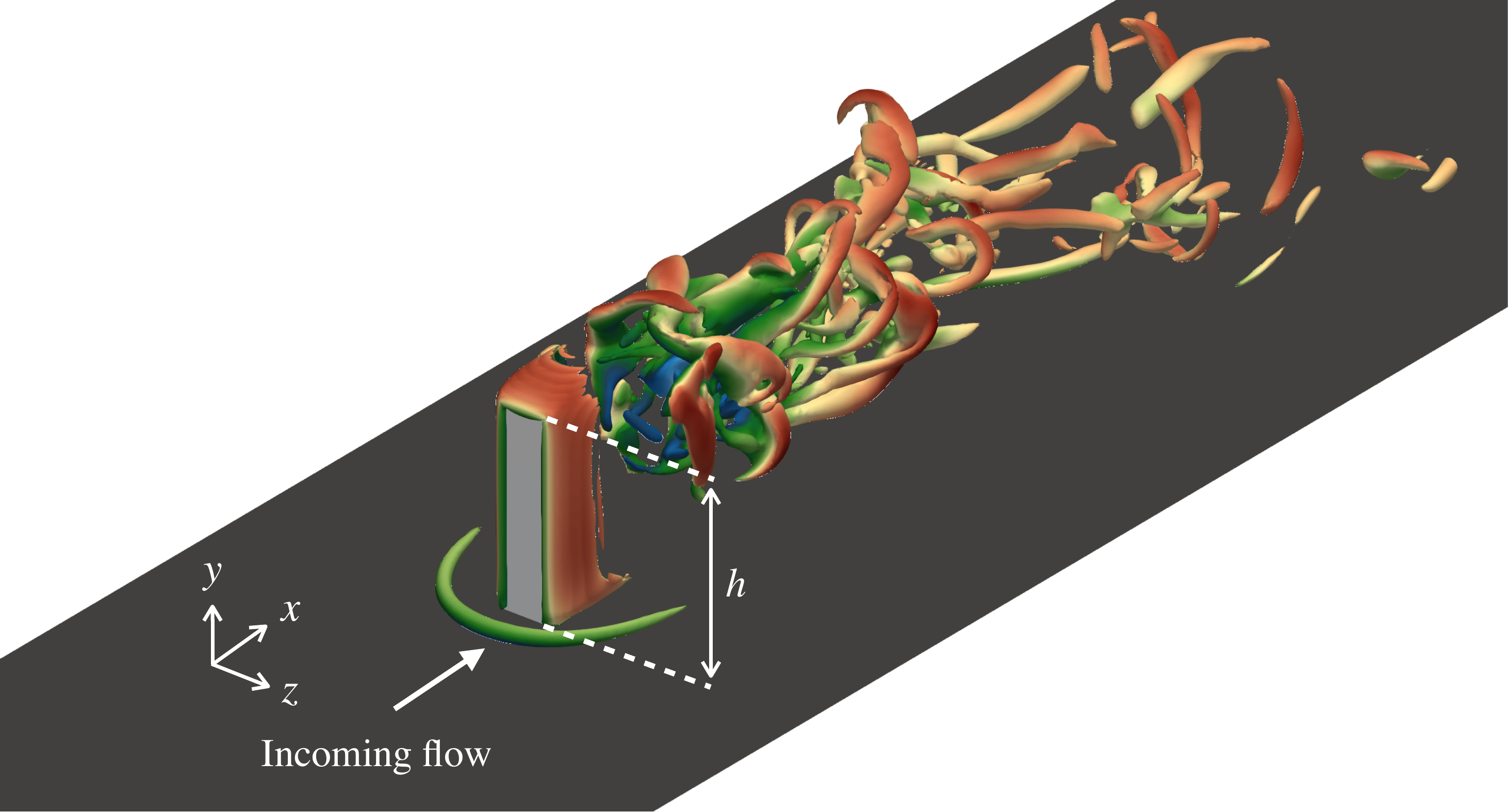}}
  \caption{Instantaneous visualisation of the flow around a wall-mounted square cylinder considered here. The instantaneous vortical structures identified with Q-criterion are shown with an isosurface of $+10$ (scaled in terms of $U_\infty$ and $h$). Structures are coloured with the streamwise velocity which ranges from $-0.79$ (dark blue) to $+1.23$ (dark red). Dark grey represents the bottom wall, whereas light grey indicates the building-like obstacle.}
\label{fig:Domain}
\end{figure}

This article presents the analysis of the flow around a wall-mounted square cylinder, following several similar studies~\citep{Torres2021,martinez2022,marco2022}, with the difference that in this case the inflow boundary layer is laminar. This database was obtained through direct numerical simulation (DNS), using the open-source numerical code Nek5000~\citep{Nek5000}, which is based on the spectral-element method (SEM) to solve the incompressible Navier--Stokes equations:
\begin{equation}
\begin{aligned}
  \nabla\cdot\boldsymbol{\rm{u}} &= 0, \\
  \frac{\delta \boldsymbol{\rm{u}}}{\delta t} + \left(\boldsymbol{\rm{u}}\cdot\nabla\right)\boldsymbol{\rm{u}} &= -\nabla p + \nu \nabla^2\boldsymbol{\rm{u}},
\end{aligned}
\end{equation}
where $\boldsymbol{\rm{u}}(x,y,z,t)$ represents the velocity field, $\nu$ is the kinematic viscosity and $p$ denotes the pressure, which includes the constant-density term. The geometrical domain comprises a single wall-mounted square cylinder, as depicted in figure \ref{fig:Domain}, with a width-to-height ratio $b/h=0.25$ and a Reynolds number based on freestream velocity and obstacle width of 500. All dimensions are normalised with the height of the obstacle, $h$, and every velocity component is normalised with the free-stream velocity, $U_\infty$. The spectral-element method combines the geometrical flexibility required to discretise the domain with the high-order accuracy of spectral methods. We consider a spectral-element mesh of $100,935$ hexadral elements with a six-point Gauss--Lobatto--Legendre (GLL) quadrature, leading to $21.8$ million grid points, to solve the scale disparity of the flow. Additional details on the numerical scheme and employed resolution can be found in a similar study~\citep{marco2022}.

As we focus on the flow near the obstacle, the following region is extracted from the computational domain: $-1\leq x/h \leq5$, $0\leq y/h \leq 2$ and $-1.5\leq z/h \leq1.5$. Using this reduced domain, we consider $10{,}000$ three-dimensional instantaneous fields to perform the modal decompositions. The previous fields were spectrally interpolated from the original SEM mesh to another one with a resolution $\left(N_x,N_y,N_z\right) = \left(300,100,150\right)$. All temporal parameters are expressed in convective time units (CTUs), \textit{i.e.} a ratio between a characteristic length and a velocity. In the present case, time is obtained from the free-stream velocity $U_\infty$ and the height of the obstacle, $h$. The time-step between snapshots is constant, ${\Delta t}_{s} = 0.005 $, which yields a database spanning a total of 50 time units. This temporal resolution is enough to accurately capture the low- and high-frequency flow mechanisms identified in the literature~\citep{martinez2022}. 

Some of the characteristics of the highly three-dimensional flow around a finite wall-mounted cylinder are displayed in the instantaneous vortical structures of figure~\ref{fig:Domain}. Here, a horseshoe vortex extends around the two sides of the obstacle and into the wake. This vortex affects many of the flow structures around the cylinder, including the vortex-shedding mode, the shear-layer dynamics and the width of the wake~\citep{rao2004,Vinuesa2015}. As a result, this vortex also has an impact on the near-wake region and, thus, on the arch vortex formed on the leeward side of the obstacle~\citep{wang2006,wang_zhou_2009}. The large range of scales characteristic of turbulent flows in urban environments is also shown in figure~\ref{fig:Domain}, where the vortical structures within the wake exhibit a wide range of sizes and energy contents. The formation of the previous vortices and wake arises as a consequence of a fixed separation location prescribed by the sharp cylinder edges, whose associated separated shear layer can be noticed around every windward edge of the obstacle.

\section{Reduced-order model for urban flows \label{sec: ROM}}

Modal decomposition is a mathematical method for identifying essential energy and dynamic characteristics of fluid flows. These spatial features of the flow are known as spatial modes and they are usually ranked in terms of the energy content levels or characteristic growth rates and frequencies driving the flow motion. These modal-decomposition techniques are generally used to create a low-dimensional coordinate system that successfully reflects the main characteristics of the flow. Not only are these structures crucial for flow analysis, but also for reduced-order modelling and flow control. In this section, we discuss the proper-orthogonal decomposition (POD), which is used to generate a low-dimensional model of the flow around a finite square cylinder, representative of a simplified urban environment.

POD~\citep{Lumley1967} is a modal-decomposition technique that has been traditionally employed in the fluid-mechanics community. It seeks to decompose a set of data for a particular field variable into the fewest feasible modes (basis functions) while yet capturing the largest amount of energy. This process implies that POD modes are optimal in minimising the mean-square error between the signal and its reconstructed representation. The low-dimensional latent space provided by the POD modes is attractive for interpreting the most energetic and dominant patterns within a given flow field. Let us consider a vector field $\boldsymbol{q}\left(\boldsymbol{\xi},t\right)$, which may represent  \textit{e.g.} the velocity or the vorticity field, depending on a spatial vector $\boldsymbol{\xi}$ and time. In fluid-flow applications, the temporal mean $\boldsymbol{\bar{q}}\left(\boldsymbol{\xi}\right)$ is usually subtracted to analyse the fluctuating component of the field variable:
\begin{equation}
    \boldsymbol x(t) = \boldsymbol{q}\left(\boldsymbol{\xi},t\right) - \boldsymbol{\bar{q}}\left(\boldsymbol{\xi}\right),\quad\quad t = t_1,t_2,\dots,t_k
\end{equation}
\noindent where $\boldsymbol x(t)$ represents the fluctuating component of the vector data with its temporal mean removed. This representation emphasises the idea that the data vector $\boldsymbol x(t)$ is being considered as a collection of snapshots at different time instants $t_k$. If the $m$ snapshots are then stacked into a matrix form, we obtain the so-called snapshot matrix $\boldsymbol X$:
\begin{equation}
    \boldsymbol X = \left[\boldsymbol x(t_1), \boldsymbol x(t_2),\dots,\boldsymbol x(t_m)\right] \in \mathbb{R}^{J\times K},
\end{equation}
\noindent where $J$ represents the number of points in $x$, $y$ and $z$. Note the similarity between this matrix and the definition provided in~(\ref{eq: Vector}). In this case, we differentiate between the snapshot matrix $\boldsymbol X\in \mathbb{R}^{J\times K}$, which is used for the modal-decomposition analysis, and the matrix $\boldsymbol{\mathcal{V}}\in \mathbb{R}^{n\times K}$, which is employed for the causality analysis. The objective of the POD analysis is to find the optimal basis to represent a given set of data $\boldsymbol x(t)$. This can be solved by finding the eigenvectors $\boldsymbol{\Phi}_j$ and the eigenvalues $\lambda_j$ from:
\begin{equation} \label{eq: POD basic eq}
    \boldsymbol C \boldsymbol{\Phi}_j = \lambda_j \boldsymbol{\Phi}_j, \quad\quad \boldsymbol{\Phi}_j \in \mathbb{R}^{J}, \quad\quad \lambda_1\geq\dots\geq\lambda_N\geq0,
\end{equation}
\noindent where $\boldsymbol C$ denotes the covariance matrix of the input data, defined as:
\begin{equation}
    \boldsymbol C = \sum_{i=1}^{K} \boldsymbol x\left(t_i\right)\boldsymbol x^\text{T}\left(t_i\right) = \boldsymbol X \boldsymbol X ^\text{T} \in \mathbb{R}^{J\times J}.
\end{equation}
The size of this matrix depends on the spatial degrees of freedom of the problem. The POD modes are derived from the eigenvectors of equation~(\ref{eq: POD basic eq}), with the eigenvalues reflecting how well each eigenvector $\boldsymbol{\Phi}_j$ represents the original data in the $\ell_2$-sense. This allows to categorise modes according to the amount of captured kinetic energy when the velocity fields are the data analysed, which enhances the assessment of the most prominent patterns in a given flow field.

\begin{figure}
  \centerline{\includegraphics[width=\textwidth]{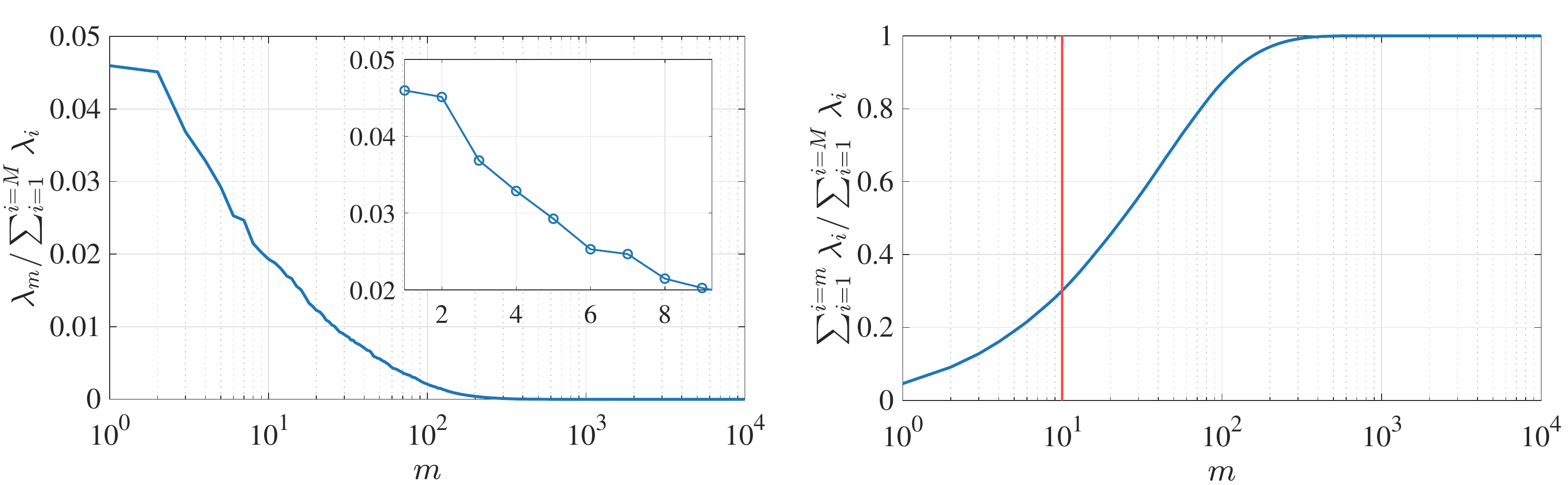}}
  \caption{Eigenvalues $\lambda_m$ (left) and cumulative sum of the eigenvalues $\sum_{i=1}^{i=m} \lambda_i$ (right) spectrum normalised with the total energy of the eigenvalues $\sum_{i=1}^M \lambda_i$. The mode number is denoted with $m$ and the solid red line represents the amount of energy contained within the first 10 modes.}
\label{fig:Spectrum}
\end{figure}

We applied POD using the singular-value-decomposition (SVD) method~\citep{Sirovich1987} on the database presented in \S\,\ref{sec: Numerical simulations}. Figure \ref{fig:Spectrum} shows the eigenvalues $\lambda_m$ and the cumulative sum of eigenvalues $\sum_{i=1}^{i=m} \lambda_i$ normalised with the total energy of the eigenvalues $\sum_{i=1}^M \lambda_i$, where $m$ is used to identify the mode number. Note that using 10 linearly-superposed modal functions, $30\%$ of the total energy can be represented, which is enough to characterise the large-scale structures driving the main dynamics in this type of flow~\citep{xiao2019,Lazpita2022}.

\begin{figure}
  \centerline{\includegraphics[width=0.48\textwidth]{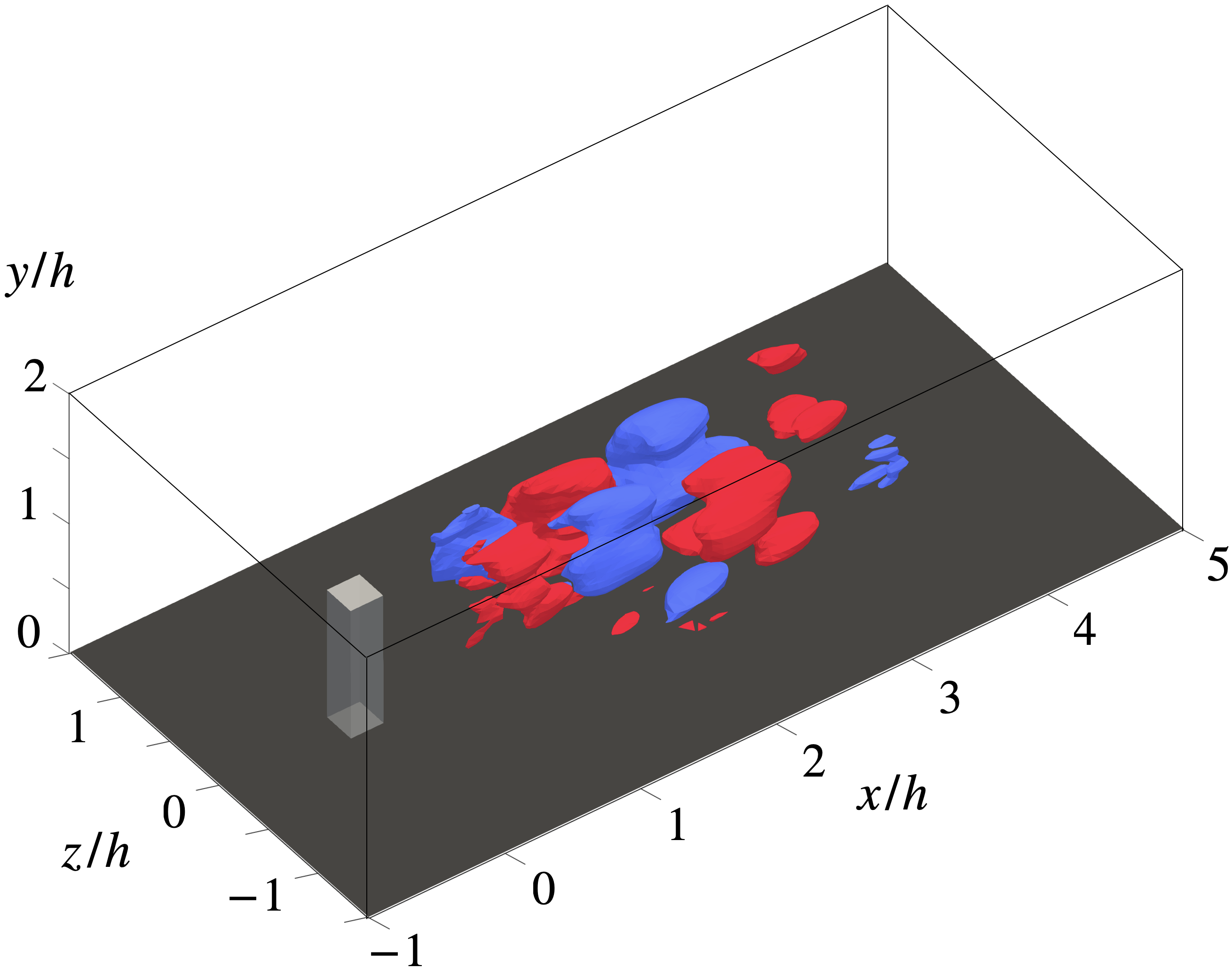}\hfill
  \includegraphics[width=0.48\textwidth]{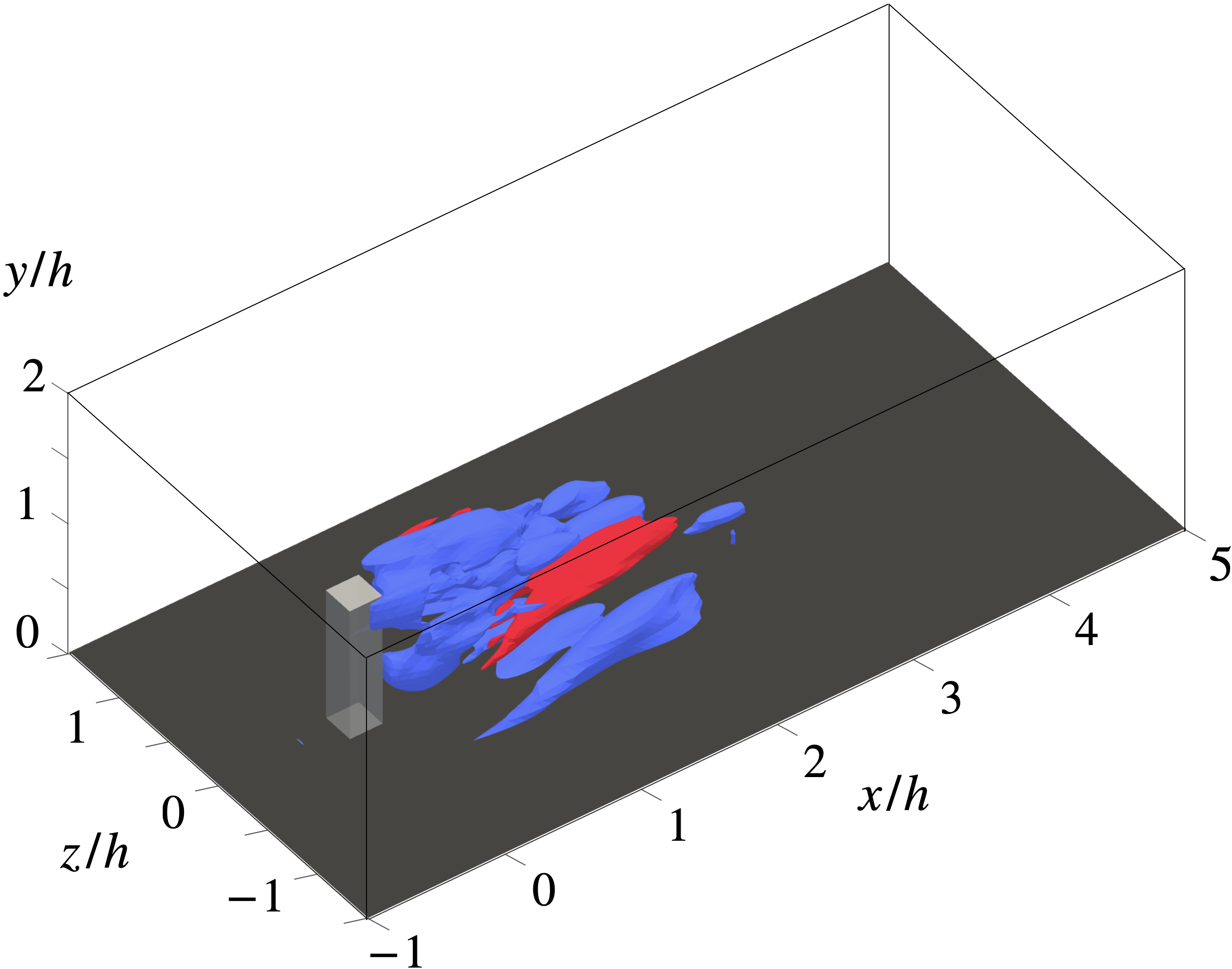}}
  \caption{Three-dimensional iso-surfaces of the streamwise velocity of the vortex-breaker (left) and vortex-generator (right) modes. Velocity values are normalised using the $L_{\infty}-$norm. Iso-values employed are given by $a\,U_{\rm{max}}$ (red) and $b\,U_{\rm{min}}$ (blue). Left: B mode with $a=b=0.3$ and right: G mode with $a=0.5$ and $b=0.1$.}
\label{fig:Modes Main}
\end{figure}

Using this ten-mode model, we base on the distinction of the flow mechanisms responsible for arch vortices in urban fluid flows performed in our previous work~\citep{Lazpita2022} to perform a similar classification. This division focuses on the identification of two main types of modes: vortex-generator modes (G) and vortex-breaker modes (B). The major structures and vortices are produced by the vortex-generator modes; therefore, they are related to the mechanism that could create the horseshoe and arch vortices. Note that G modes typically exhibit a smaller energy content than B modes and are found in the low-frequency region of the spectrum. The major flow structures could then be broken by the vortex-breaker modes, which are also responsible for the dynamics of the turbulent wake. B modes, in contrast to vortex-generators, are present in the high-frequency region of the spectrum.

Since regions of strong recirculation have already been proved to increase concentration of passive scalars~\citep{zhu2017}, G modes, which are related to these prominent recirculation areas, could be related to high-pollutant-concentration areas~\citep{Monnier2018}. Conversely, vortex-breaker modes could be connected with reduced-pollutant-concentration regions as they are responsible for the destruction mechanisms of the main spatio-temporal structures. The readers are referred to \cite{Lazpita2022} and \cite{martinez2022} for a detailed discussion of the previous modes. 

\begin{figure}
  \centerline{\includegraphics[width=\textwidth]{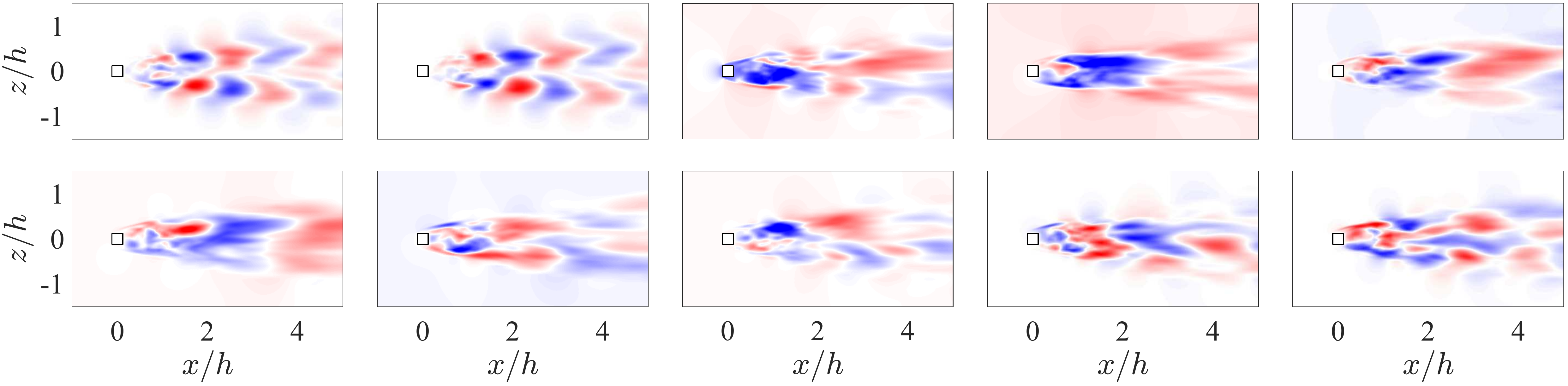}}
  \centerline{\includegraphics[width=\textwidth]{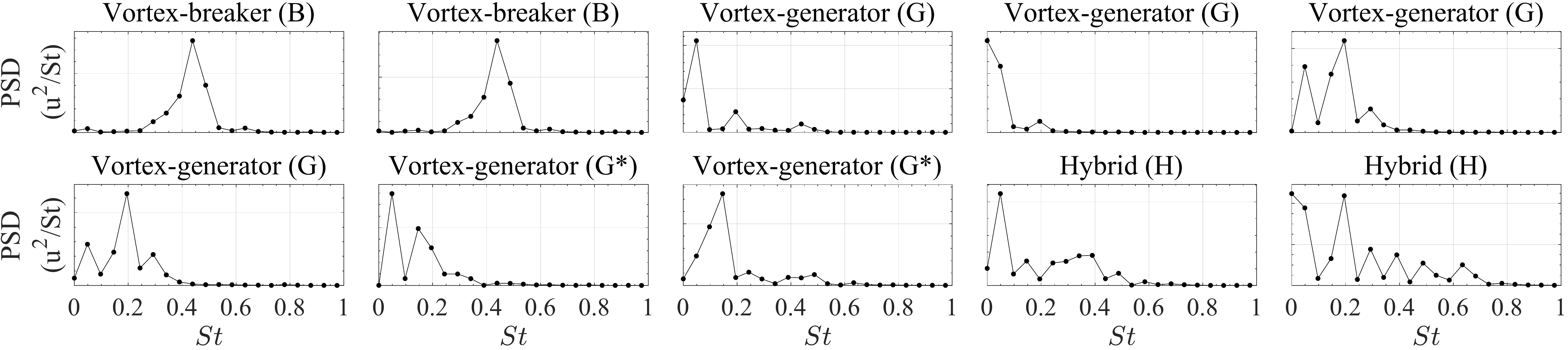}}
  \caption{(From upper left to lower right on two top panels) First ten POD modes at $y/h=0.75$ for the streamwise component of the velocity. Contours are normalised with the $L_\infty$-norm and range between $-1$ (blue) and $+1$ (red). (Bottom) Power-spectral density (PSD) scaled with the Strouhal number $St=f h/U_\infty$ of the temporal coefficients associated with the corresponding POD modes, where $f$ is the characteristic frequency of each mode. The PSD is calculated using $N=8192\,(2^{13})$ samples and a window overlap of $50\%$.}
\label{fig:Modes 10}
\end{figure}

The three-dimensional structures characteristic of these two types of modes are represented in figure~\ref{fig:Modes Main} for the present database. The vortex-breaker mode is shown in figure~\ref{fig:Modes Main} (left). This type of mode is represented by the first two POD modes shown in figure~\ref{fig:Spectrum}. Therefore, the vortex-breaking process is identified as the most energetically-relevant mode present in the flow field. This result is in line with \cite{Lazpita2022}, who showed the agreement between the first two POD modes and the vortex-breaker modes identified using various modal-decomposition techniques, both in terms of frequency behaviour and spatial resemblance. Here, the streamwise turbulent wake consists of high-velocity coherent clusters on both sides of the obstacle wake, which are also related to the vortex-shedding phenomenon present in the flow past bluff bodies. This is also shown in the two-dimensional contours in figure~\ref{fig:Modes 10}, where the spatial structure of the two modes is observed to be the same, except for a shift in phase: they are both antisymmetric with respect to the $z$-axis and they both represent the vortex-shedding phenomenon with identical frequency, $St = 0.45$. This is an evidence that the modes represent a wave-like periodic structure of the flow. In fact, since the POD modes are real, two modes are needed to describe a flow structure travelling as a wave~\citep{rempfer1994}.

A similar analysis can be conducted using the third POD mode, whose associated structures are representative of G modes. In figure \ref{fig:Modes Main} (right), we show that this mode is connected with the main vortex-generation mechanisms identified in \cite{Lazpita2022}. A large-scale streamwise structure, characteristic of this type of mode, is found just after the obstacle. This dome-like feature surrounds and interacts with the arch vortex by restricting its expansion~\citep{martinez2022}. 

The rest of the modes are depicted in figure~\ref{fig:Modes 10}, where the spatial modes for the streamwise velocity fields are shown. An additional analysis of the temporal coefficients associated with these modes is conducted in the frequency domain through the fast-Fourier-Transform (FFT) method~\citep{FFT}. This enables classifying the time coefficients associated with each spatial mode into low- and high-frequency phenomena, the features of which are decisive for the vortex-generating and breaking processes, respectively. Remarkably, these results prove that the dynamics of the flow are driven by the vortex-breaking process present in the first two most-energetic modes. These modes are dominated by a peak frequency of $St = 0.45$. Modes 3 and 4 are denoted as vortex-generator modes since their associated peak frequencies appear in the low-frequency region of the domain ($St =0.05$ and $0$, respectively) and their associated spatial structures, depicted in figures~\ref{fig:Modes Main} and~\ref{fig:Modes 10}, show how a dome-like structure encloses the near-wake region of the flow. This behaviour is similar to that of the time-averaged field, thus it suggests that these modes could be the reason for the generation of such structures. Higher-order modes, \textit{i.e.} modes 5 to 8, are also present in the low-frequency range of the spectrum. However, their flow features start to exhibit fluctuating features in the wake. These modes may then be regarded as vortex-generator modes that are harmonics of the previous G-modes with dominant frequencies $St=0.05$, $0.15$ and $0.2$. However, these modes are the result of nonlinear interactions; hence, they can be regarded as hybrid modes (which is the case of modes 9 and 10), whose interaction with each other is expected to be extracted from the causality analysis.


\section{Results and discussion} \label{sec: Results}

\begin{figure}
  \centerline{\includegraphics[width=0.6\textwidth]{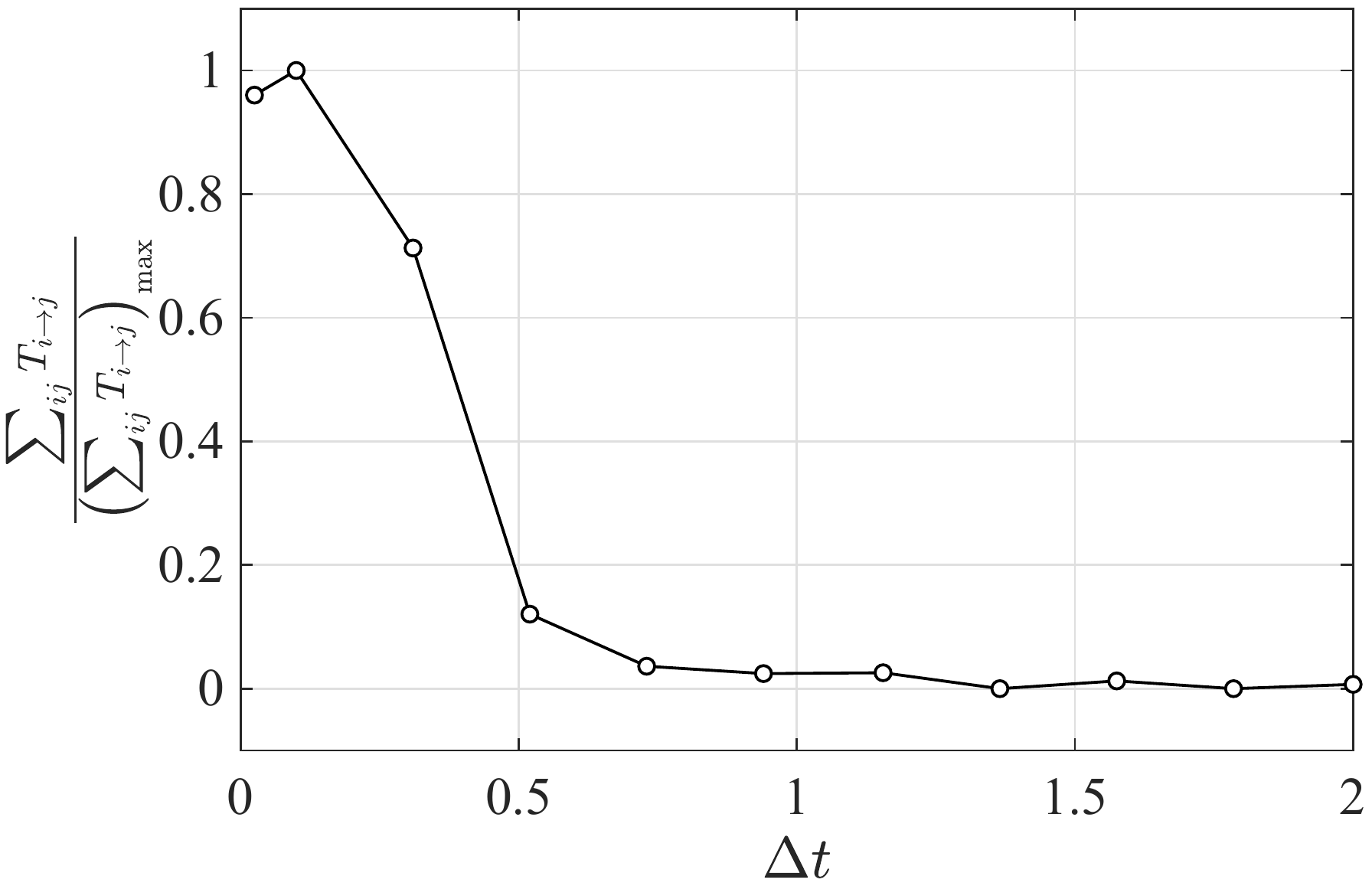}}
  \caption{Evolution of total causality $\sum_{ij} T_{i\rightarrow j}$ as a function of time lag $\Delta t$. Note that $\Delta t$ is scaled in terms of the time-step between snapshots, ${\Delta t}_s$, and total causality is normalised with the maximum value obtained for every $\Delta t$, \textit{i.e.} $\left(\sum_{ij} T_{i\rightarrow j}\right)_{\rm{max}}=0.031$.}
\label{fig:Causality_vs_dt}
\end{figure}

Using the previous ten-mode model, we apply the procedure discussed in \S\,\ref{sec: Moehlis} to extract the causal interactions between the different flow mechanisms. To do that, we use the time coefficients associated with each of the POD modes and we arrange them into a ten-component vector as in (\ref{eq: Vector}). Prior to the assessment of the causal results, the quantification of causality using (\ref{eq: Transfer Entropy Vector}) requires the definition of a certain fixed-time delay, $\Delta t$. In this study, we seek for the time-lag that produces the maximum causal inference between the variables, which we denote as ${\Delta t}_{\rm{max}}$. Despite the inherent change in transfer-entropy behaviour for varying $\Delta t$ values, defining the summation of every causal relationship as a global measure, \textit{i.e.} $\sum_{i,j} T_{i\rightarrow j}$, allows to establish a sensible value for ${\Delta t}_{\rm{max}}$. The evolution of the previous parameter is depicted in figure~\ref{fig:Causality_vs_dt} as a function of the time lag, where ${\Delta t}$ is scaled in terms of the time-step between snapshots ${\Delta t}_s$. Causalities are found to be maximum for a time lag ${\Delta t}_{\rm{max}} = 0.1$, which corresponds to a lag of $20$ snapshots. This value is comparable to the dynamics of the highest-frequency phenomena of the flow~\citep{martinez2022} and represents $5\%$ of the frequency of the vortex-shedding modes.

\begin{figure}
  \centerline{\includegraphics[width=0.49\textwidth]{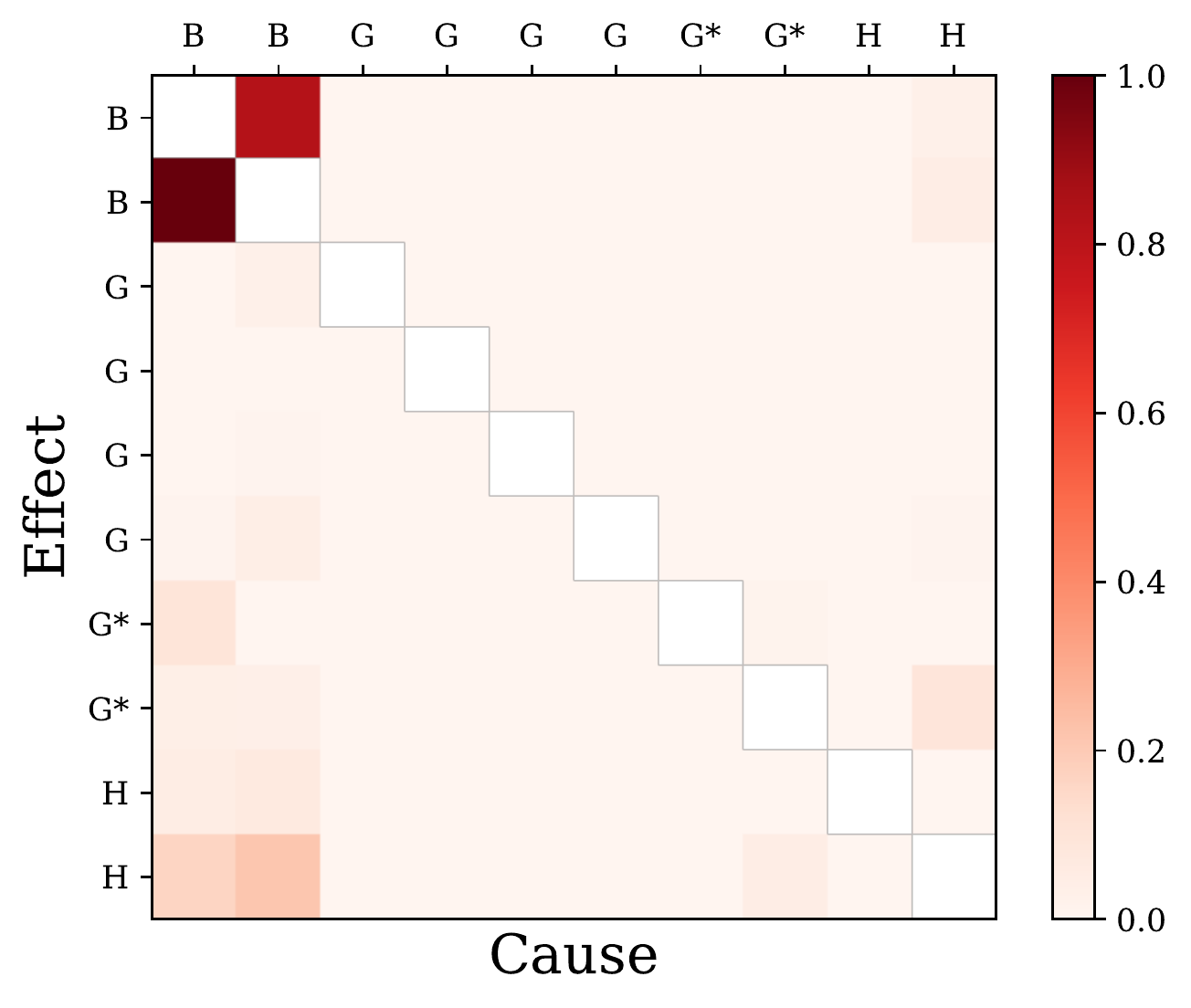}
  \hfill
  \includegraphics[width=0.49\textwidth]{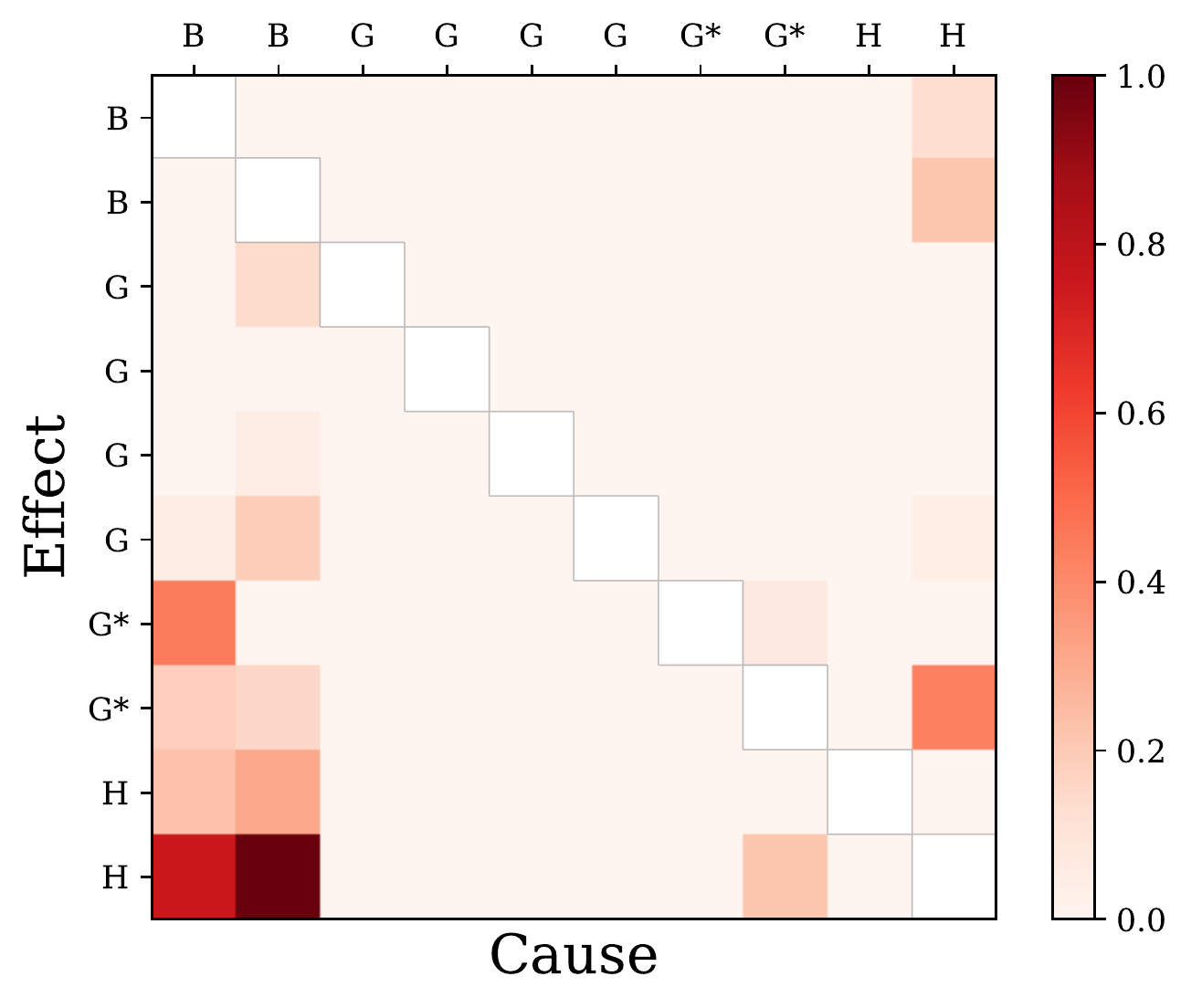}}
  \caption{Causal map for the ten-mode model of the studied database. Redscale colours denote causality magnitude normalised with $L_\infty$-norm. Modes are labelled per their mechanism. The contribution of B-modes is set to zero in the right panel to highlight the causal interaction with higher-order modes.}
\label{fig:UrbanCausality}
\end{figure}

\begin{figure}
  \centerline{\includegraphics[width=0.49\textwidth]{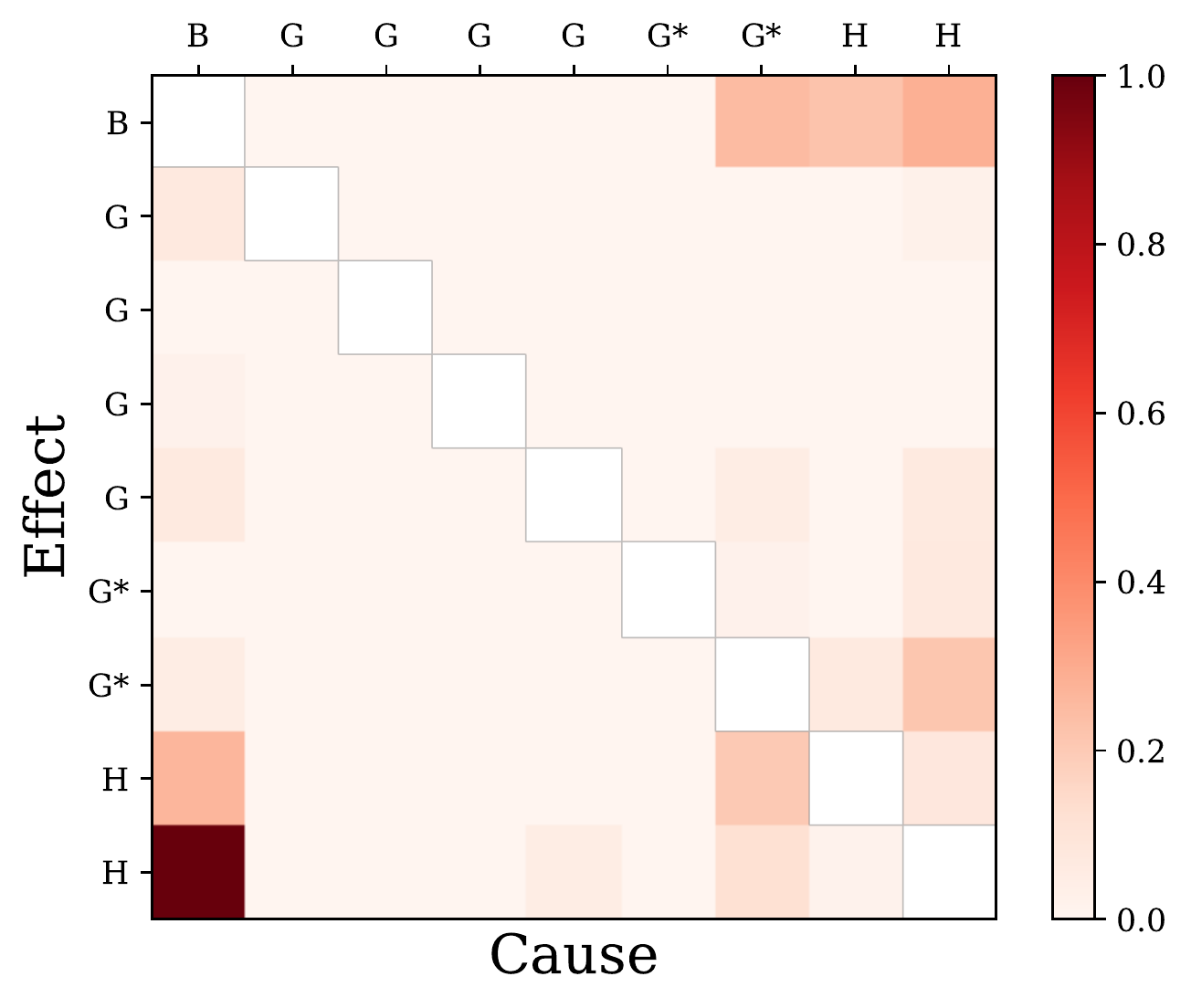}
  \hfill
  \includegraphics[width=0.49\textwidth]{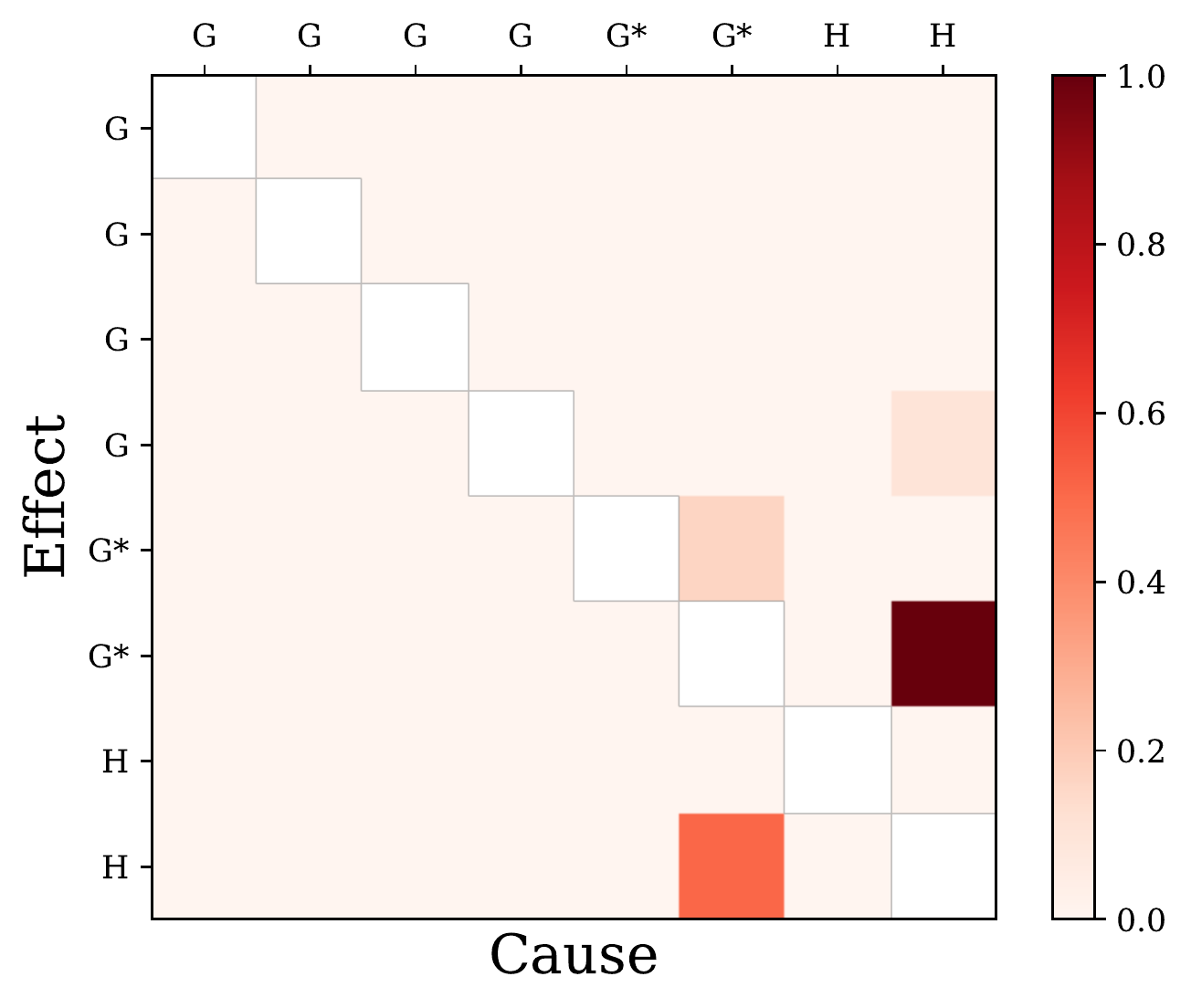}}
  \caption{Causal map for the ten-mode model of the studied database for (left) only one B-mode and (right) no B-modes. Redscale colours denote causality magnitude normalised with $L_\infty$-norm. Modes are labelled per their mechanism.}
\label{fig:UrbanCausality_2}
\end{figure}

The transfer entropy is then calculated using (\ref{eq: Transfer Entropy Vector}) with ${\Delta t}_{\rm{max}}$ to obtain the cross-induced cause-and-effect interactions between the modes. The corresponding causal map is depicted in figure~\ref{fig:UrbanCausality}, where modes are directly labelled using the flow mechanism that they represent according to \S\,\ref{sec: ROM}. The map shows a very high causal inference between the first two modes compared to the rest of relationships. The fact that these two modes are representative of the same flow mechanism but with a slight shift in phase makes their causal relationship evident. Causal inferences smaller in amplitude are also appreciated between B-modes and the highest-order modes, \textit{i.e.} modes 7 to 10. To better assess these interactions, the causal relationships between B-modes are set to zero. The resulting causal map is illustrated in figure~\ref{fig:UrbanCausality} (right), where we observe the high causal connections $T_{2\rightarrow 10}$ and $T_{1\rightarrow 10}$. Vortex-breaker modes are then regarded as the most influential mechanism in higher-order modes, which we denoted as hybrid modes. However, vortex-generator modes (modes 3 to 6) do not have a direct causal inference over the rest of the modes even though it could have been hypothesised \textit{a priori} that G-modes should occur before B-modes.

Similar conclusions can be drawn if we decide to consider only one vortex-breaker mode, since these modes represent the same flow mechanism and their associated temporal evolution is equivalent but with a $\pi /2$ phase shift. This mechanism propagates as a travelling wave in a periodic fashion; therefore, two orthogonal modes in space are needed to represent its dynamics. This leads to the casual map shown in figure~\ref{fig:UrbanCausality_2} (left), where only one B-mode can be noticed. This assumption produces effects similar to those found from modes 1 and 2 on higher-order modes. As seen in \S\,\ref{sec: ROM}, hybrid modes represent the interaction between vortex-breaker and generator modes as a result of a varied range of frequencies found in the spectrum of their associated time coefficients. Therefore, $T_{B\rightarrow 10}$ reveals that the vortex-breaking mechanism drives the appearance of modes where shared features of both B- and G-modes are found.

\begin{figure}
  \centerline{\includegraphics[width=\textwidth]{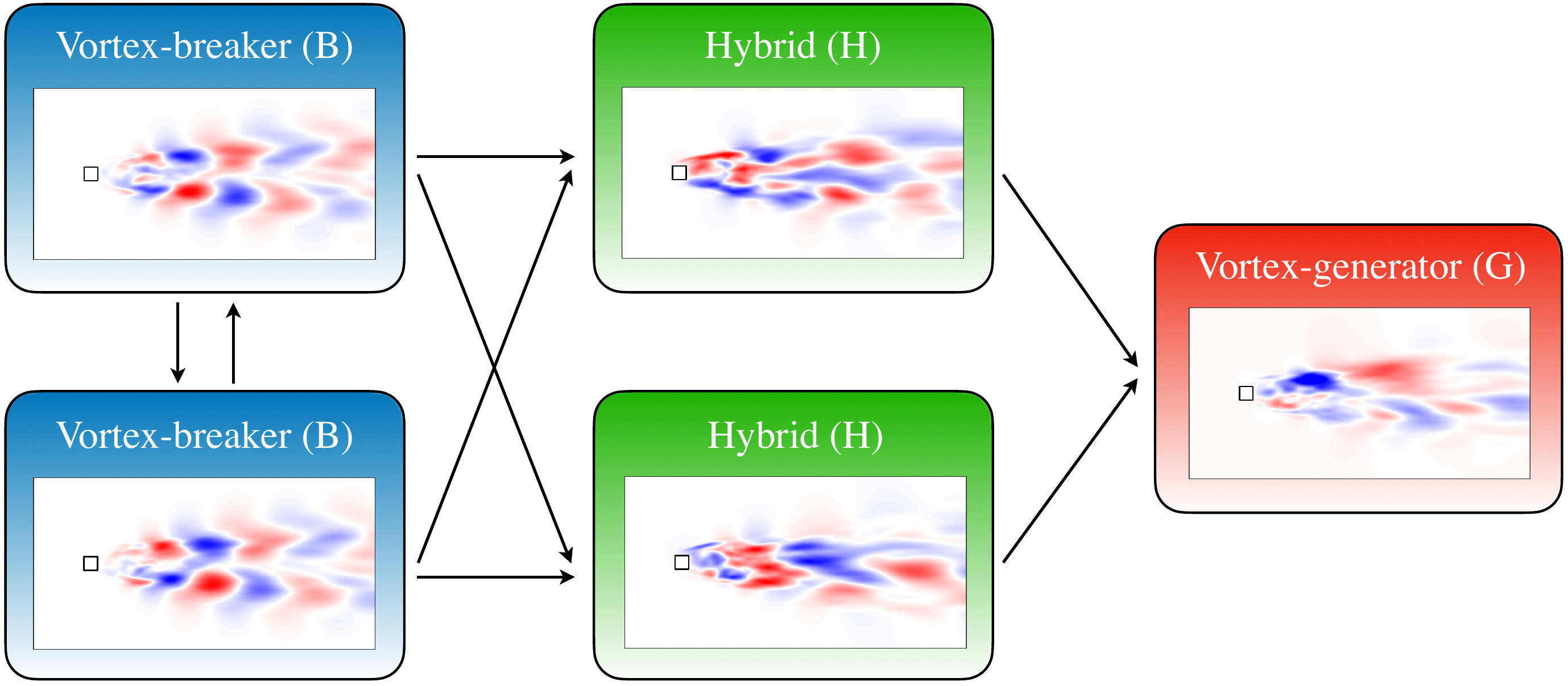}}
  \caption{Diagram depicting the mutual inferences between each mode due to the past states of the other modes. Only those significant causal interactions are represented. Two vortex-breaker (B), two hybrid (H) and one vortex-generator (G) modes are shown, \textit{i.e.} modes 1, 2, 10, 9 and 8, respectively.}
\label{fig:UrbanCausality_Diagram}
\end{figure}

Figure~\ref{fig:UrbanCausality_2} (right) depicts the eight-mode causal map obtained as a result of removing from the spectrum the combined mode. Remarkably, a strong causal inference from mode 10 on 8, $T_{10\rightarrow 8}$, is now appreciated. This proves that modes on which B-modes have large influence also cause the appearance of modes whose frequency behaviour is harmonic of purely G-modes. However, there is no sign that vortex-generator modes are intimately related to any of the aforementioned mechanisms. A schematic diagram depicting the main causal interactions between each of the modes is presented in figure~\ref{fig:UrbanCausality_Diagram}.

Therefore, understanding the dynamics underlying the vortex-breaking mechanism in detail may be crucial for predicting the emergence of significant flow features in urban environments, such as the arch vortex. These findings also suggest that B-modes are not only the most energetic modes in the flow field but also those responsible for the appearance of modes that relate to the generation of the primary vortical structures.

\subsection{Time correlation between time coefficients}
In this section, we compare the previous results with those obtained using a simplified approach for the quantification of causality, \textit{i.e.} time correlation. This statistical metric describes the magnitude of the relationship between a given pair of variables without the directionality and asymmetry properties, which are required to estimate the causes and effects of events. Additionally, high correlation between the variables does not automatically mean that changes in one variable are caused by changes in the other variable. Despite this, it is interesting to compare the results from both approaches since a high causation emerges from some degree of correlation between variables, although the opposite does not hold. To assess correlation, the following expresion is used:
\begin{equation}
    C_{ij}\left(\Delta t\right) = \frac{ \langle{{\mathcal{V}_i}'\left(t\right),{\mathcal{V}_j}'\left(t+\Delta t\right)}\rangle}
    {\norm{{\mathcal{V}_i}'\left(t\right)}\cdot\norm{{\mathcal{V}_j}'\left(t\right)}},
\end{equation}
where ${\mathcal{V}_i}'$ and ${\mathcal{V}_j}'$ represent the fluctuating signals ${\mathcal{V}_i}'={\mathcal{V}_i}-{\overline{\mathcal{V}}_i}$ and $\langle\cdot\rangle$ denotes the dot product operation taken over the whole time history. 

\begin{figure}
  \centerline{\includegraphics[width=0.6\textwidth]{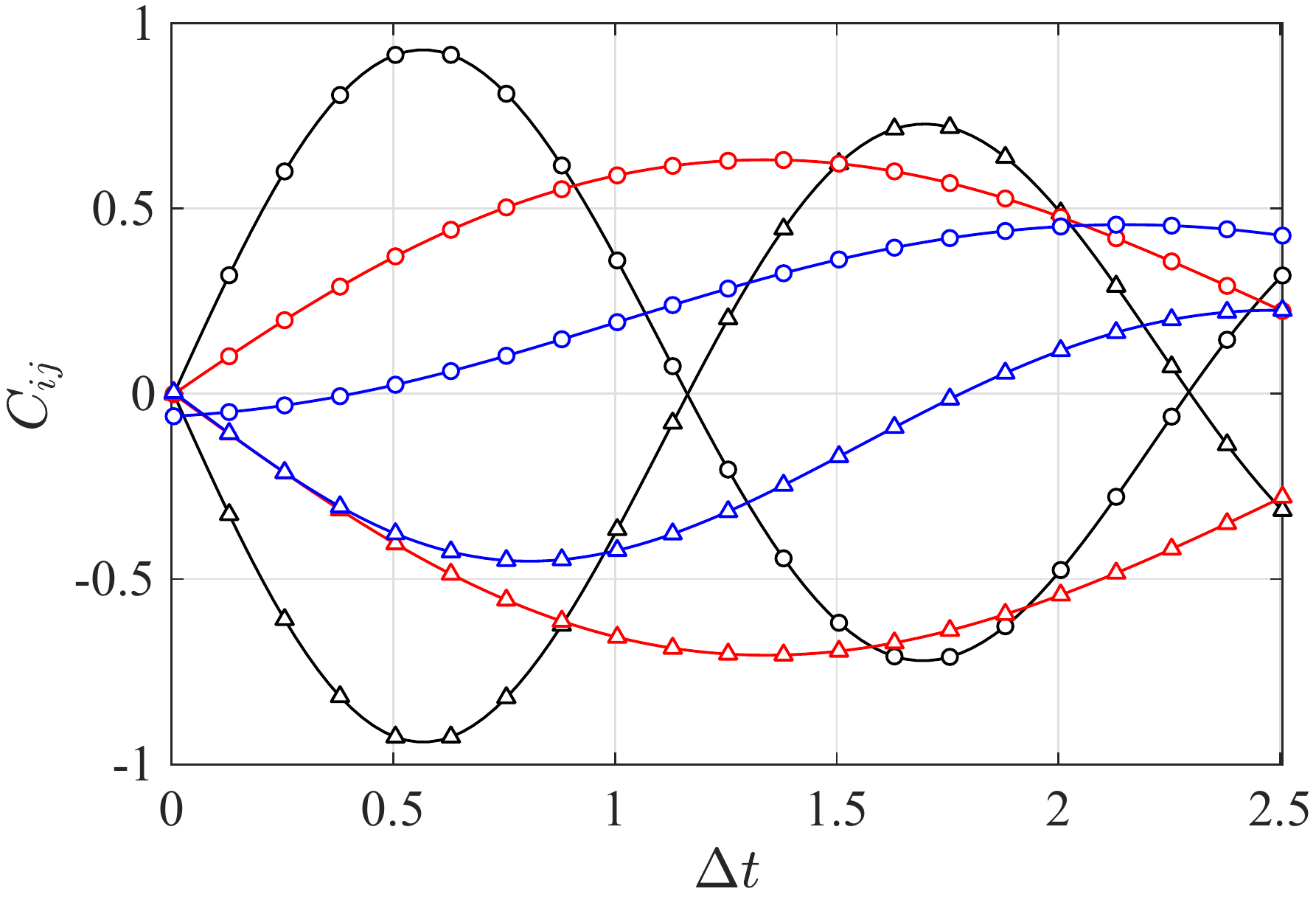}}
  \caption{Time correlation between $a_1\rightarrow a_2$, $\circ$; $a_2\rightarrow a_1$, $\footnotesize{\bigtriangleup}$; $a_5\rightarrow a_6$, $\color{red}\circ$; $a_6\rightarrow a_5$, {$\color{red}\bigtriangleup$}; $a_7\rightarrow a_5$, {$\color{blue}\circ$} and $a_9\rightarrow a_8$, {$\color{blue}\bigtriangleup$}, where $a_m$ refers to the $m$-th mode of the reduced-order model.}
\label{fig:Time_Correlation}
\end{figure}

Figure~\ref{fig:Time_Correlation} depicts the effects of the variation of the time lag on the temporal cross-correlation between the time-coefficients signals. Only those trends with a maximum value above $0.4$ are represented. The main conclusion from these results is the high correlation observed between the first two modes. Remarkably, the correlation approaches the value of 1 when ${\Delta t}/T_{1,2} = 0.25$, where $T_{1,2}$ represents the period of the oscillations of the corresponding signals. This evolution confirms that the previous modes are representative of the same flow mechanism with a phase shift of $\pi /2$ rad. This is in line with the causal relations observed in the previous section, where these modes represented the largest causation. A similar trend is also observed for modes 5 and 6, both G-modes. In particular, a high correlation is observed when $C_{1,2} = C_{2,1} = 0$. As the activity of the correlation between vortex-breaker modes decays, the associated flow structure transitions to higher-order modes. This results in an increased interaction activity of modes 5 and 6, as correlation between modes 1 and 2 decays.

Furthermore, every signal discussed in this section exhibits a damped sinusoidal behaviour with a frequency pattern similar to that observed in figure~\ref{fig:Modes 10}: B-modes are associated with high-frequency oscillations, whereas G-modes are dominated by low-frequency ones. This means that depending on the employed time-lag, the correlation between variables becomes reversed, which can be related to the fact that POD modes are based on a linear relationship. A similar conclusion was drawn from the transfer-entropy approach, where the modified values of the time lag produced different cause-and-effect interactions specially between the first two modes. Besides, damped waves are representative of poor correlations for increased values of the time-lag, which hinders the extraction of both correlation and causation trends between variables.

\section{Conclusions and further discussion} \label{sec: Conclusions}

In the present work, we analysed the formation mechanisms of large-scale coherent structures in the flow around a wall-mounted square cylinder. We employed a database obtained by means of a direct numerical simulation to solve the flow around a single finite square cylinder of width-to-height ratio $b/h=0.25$. Proper-orthogonal decomposition was then applied to the previous database to generate a reduced-order model with the ten most energetic modes, which represent $30\%$ of the overall energy of the flow. This model enables isolating the main mechanisms driving the dynamics of the flow in terms of space and time. The previous modes were classified as vortex-generator (G) and vortex-breaker (B) modes as a result of their spatial features and associated frequency behaviour. We investigated the causal interactions between these mechanisms with the objective of understanding the origin and evolution of the various three-dimensional topological patterns that precede the formation of the well-known vortical structures found in this type of flows. To that end, we applied conditional transfer entropy over the time coefficients associated to each of the POD modes. This approach is based on a information-theoretical quantity that quantifies the amount of information flowing from the past state of one variable to another. In particular, we identified B modes as the most causative modes over higher-order modes, which we denoted as hybrid (H) modes due to their shared features with B and G modes. The cases in which B modes have large influence, \textit{i.e.} H modes, were also found to drive the appearance of modes whose frequency behaviour is harmonic of purely G modes. Besides, no significant causal relationships were observed for G modes, a fact that highlights the importance of understanding the underlying dynamics of B mechanisms to predict the emergence of significant flow features in urban environments. These results go in line with the previous classification of modes and shed light on new possibilities for future urban-flow control research. This includes active flow control targeting the modes responsible for generation of large-scale structures which may increase pollutant concentration in cities. 

The tool employed for the quantification of causality was previously validated through the analysis of the causal relations present in a low-dimensional model for turbulent shear flows developed by \cite{moehlis2004}. We identified three main causal relations, which were related to the lift-up mechanism, the generation of streamwise rolls and the mean-flow instability in the spanwise direction. All of them were in accordance with the results reported by \cite{lozano2020} for turbulent channel flows. 

We conclude our work with a brief discussion over the main limitations of the methods employed here. Firstly, a reduced-order model of ten linearly-superposed modes is accurate enough to represent the large-scale structures of the flow. However, these results remain to be confirmed with other models containing a larger number of modes, in which other structures not included in the previous ROM might have an impact on the overall causal inferences. Other modal-decomposition techniques could also be explored such that the dynamics of the global flow field are represented more efficiently. Moreover, the identification of causal interactions between variables remains an ongoing challenge. Appropriate metrics have to be defined, especially given the vast number of parameters whose selection is critical in the resulting causality. In particular, typical operationalisations of causality estimators are based on a strong assumption that each point in the time series of effects is influenced by a combination of other time series with a fixed time lag. This is usually not the case in real-life systems; therefore, a clear generalisation is needed to relax the above assumption. More importantly, the notion of causality followed here is based on Shannon statistical entropy and should therefore be interpreted as a probabilistic measure of causality and not as the measurements of individual events. 

\section*{Acknowledgements}
R.V. acknowledges the financial support by the G\"oran Gustafsson and the Lundeqvist foundations. A.S. acknowledges support from the NSF CAREER, USA grant No. 1554033 to the Illinois Institute of Technology.
S.L.C. acknowledges the grant PID2020-114173RB-I00 funded by MCIN/AEI/10.13039/501100011033.
E.L. acknowledges Thomas and Josette Morel for their generosity, support, and the creation of the Morel Fellowship which allowed this work to occur.

\section*{Declaration of interests}
The authors report no conflict of interest.

\bibliographystyle{jfm}
\bibliography{jfm-main}

\end{document}